\documentclass[runningheads]{llncs}
\usepackage[T1]{fontenc}
\usepackage{graphicx}
\usepackage{subcaption}
\begin{document}
%
% typical ICDAR title
%\title{Estimation of Scores and Grades Based Only on Log-Normal Parameters of Drill Handwriting Answers}
%\title{From Online Handwriting Motor Parameters to Scores and Grades:  A Sigma Log-normal Modeling Approach}
% typical HCI title
%\title{Are Online Handwriting Motor Parameters Sufficient to Estimate Scores and Grades?}
\title{Prediction of Grade, Gender, and Academic Performance of Children and Teenagers from Handwriting Using the Sigma-Lognormal Model}

\titlerunning{Grade, Gender, and Academic Performance predictions}

% If the paper title is too long for the running head, you can set
% an abbreviated paper title here
%
\author{Adrian Iste\inst{1} %\orcidID{0000-1111-2222-3333} 
\and
Kazuki Nishizawa\inst{1} %\orcidID{1111-2222-3333-4444} 
\and
Chisa Tanaka\inst{1}%\orcidID{2222--3333-4444-5555} 
\and
Andrew Vargo\inst{1}%\orcidID{2222--3333-4444-5555} 
\and
Anna Scius-Bertrand\inst{2}%\orcidID{2222--3333-4444-5555} 
\and
Andreas Fischer\inst{3}
\and
Koichi Kise\inst{1,4}%\orcidID{2222--3333-4444-5555}
}
\authorrunning{A Iste et al.}
% First names are abbreviated in the running head.
% If there are more than two authors, 'et al.' is used.
%
\institute{Osaka Metropolitan University, Osaka, Japan \and
University of Applied Sciences and Arts Western Switzerland, Fribourg, Switzerland\\
%\email{lncs@springer.com}
%\url{http://www.springer.com/gp/computer-science/lncs} 
\and
University of Fribourg, Fribourg, Switzerland \\
%\email{\{abc,lncs\}@uni-heidelberg.de} 
\and
German Research Center for Artificial Intelligence, Japan Laboratory, Osaka, Japan}
\maketitle              % typeset the header of the contribution
\begin{abstract}
Digital handwriting acquisition enables the capture of detailed temporal and kinematic signals reflecting the motor processes underlying writing behavior. While handwriting analysis has been extensively explored in clinical or adult populations, its potential for studying developmental and educational characteristics in children remains less investigated. In this work, we examine whether handwriting dynamics encode information related to student characteristics using a large-scale online dataset collected from Japanese students from elementary school to junior high school. We systematically compare three families of handwriting-derived features: basic statistical descriptors of kinematic signals, entropy-based measures of variability, and parameters obtained from the sigma-lognormal model. Although the dataset contains dense stroke-level recordings, features are aggregated at the student level to enable a controlled comparison between representations. These features are evaluated across three prediction tasks: grade prediction, gender classification, and academic performance classification, using Linear or Logistic Regression and Random Forest models under consistent experimental settings.
The results show that handwriting dynamics contain measurable signals related to developmental stage and individual differences, especially for the grade prediction task. These findings highlight the potential of kinematic handwriting analysis and confirm that through their development, children's handwriting evolves toward a lognormal motor organization.

\keywords{Online Handwriting \and Grade Prediction \and Gender Classification \and Academic Performance Classification \and Sigma-Lognormal Model \and Young Students.}
\end{abstract} %Adrian [1 page incl. title, authors, affliations]
\section{Introduction} %Adrian [1 page]

Handwriting is a complex motor activity resulting from the interaction between cognitive, perceptual, and neuromotor processes \cite{plam2013}. With the widespread adoption of digitizing devices, online handwriting acquisition now enables the capture of fine-grained temporal and kinematic signals, including pen velocity, pressure, tilt, and spatial trajectories. These measurements provide access to the dynamic aspects of writing behavior and offer new opportunities to study motor execution beyond the visual appearance of written traces. Unlike offline handwritten images, online recordings preserve the temporal structure of strokes and provide insight into the underlying motor control processes that generate written movements.

Beyond character recognition, online handwriting data have been extensively utilized for authentication and biometric verification, writer identification, and the evaluation of writing skill and motor proficiency. These data have also facilitated developmental and aging studies by capturing the evolution of motor execution across the lifespan, as well as neurological and clinical assessments of disorders affecting motor control. Among modeling approaches, the sigma-lognormal model \cite{oreilly2009sigma} offers a principled neuromotor framework that describes handwriting movements as the superposition of lognormal components associated with motor impulses. This model has demonstrated relevance in clinical contexts, including the assessment of neuromotor impairments such as Alzheimer’s disease \cite{ales}.

Despite these advances, several potentially informative tasks remain insufficiently explored, especially in younger populations. Most prior studies have focused on adult writers or specific pathological conditions. In contrast, handwriting during childhood reflects ongoing motor learning, cognitive maturation, and progressive automation of movement patterns. Because children’s handwriting is still under development, it exhibits higher variability, less stable motor coordination, and evolving control strategies. As a result, tasks such as age prediction, gender classification, or academic performance classification may be particularly challenging in school-aged populations.

In this work, we address these issues using a large-scale online handwriting dataset collected with Wacom tablets from Japanese students aged 7 to 15 years. The dataset includes approximately fifty writers per age group and around 110 drills per student. Rather than relying on computer vision approaches based on static images, we focus exclusively on features derived from writing movements themselves. We compare three families of representations: basic descriptors of kinematic signals such as speed and acceleration, entropy-based measures capturing variability and complexity in writing dynamics, and parameters derived from the sigma-lognormal model, which represents handwriting movements as the superposition of lognormal velocity primitives reflecting underlying motor commands. Although the recordings are available at the stroke level, we consider aggregated student-level representations in order to establish a first systematic comparison across feature types under consistent experimental conditions.

We evaluate these feature sets across three prediction tasks reflecting different aspects of individual variability: grade prediction, gender classification, and academic performance classification, using linear models and random forest algorithms.

The contributions of this study are as follows : we provide an analysis of online handwriting dynamics in a school-aged population, a demographic that remains underrepresented in prior work. Also, we investigate whether motor-oriented handwriting representations can serve as complementary indicators of developmental stage and academic-related characteristics. Through this analysis, we aim to better understand the potential of handwriting dynamics as a tool for studying development and learning processes in educational contexts, especially for young students.

\section{Related Work} %Adrian [1.5 pages]

The prediction of age and gender based on handwriting has been explored by various researchers. Alaei et al. offer a comprehensive review of existing methodologies, emphasizing the diversity of approaches, which range from traditional statistical analyses to advanced deep learning frameworks for these tasks \cite{ala2023}.

Handwriting analysis has been approached through both offline and online datasets. For age prediction from handwriting, existing studies primarily address multi-class classification problems, typically distinguishing between “children” and “adults”. To the best of our knowledge, no research has focused on predicting features for children under 11 years of age.

For age prediction, offline handwriting has been widely used : Basavaraja et al. used an offline dataset to predict the age of writers across four classes, with ages ranging from 11 to 24 years old \cite{basa2019}. They also employed the IAM dataset, which includes ages from 25 to 56 years, and the KHATT dataset, covering ages from 16 to 56 years.
AL-Qawasmeh et al . created an offline handwriting dataset called FSHS comprising approximately 2000 writers aged from 15 to 75 \cite{naj2022}. They extracted features from the samples using CNN and attempted to predict the age and gender of the writers. For grade prediction, they proposed classifying writers into two categories: “young adult writer” and “mature adult writer”.
Shin et al. employed an online dataset to predict the age class of writers, distinguishing between “child” (ages 12 to 13) and “adult” (ages 19 to 59), with approximately 80 participants. Features such as pen pressure, $x$ and $y$ coordinates of the pen, and pen angle were utilized \cite{shin2022}. 
Nunez et al. extracted entropy features from an online database (IRONOFF) consisting of approximately 800 participants aged 11 to 77, divided into six different classes. Nevertheless, since the class were unbalanced, they artificially reduced the number of writers to 144 \cite{rosa2016}. 

For gender classification, most studies have been made using offline handwriting datasets : 
Maken et al. utilized the Kaggle offline dataset from the ICDAR 2013 gender prediction competition, which includes approximately 280 writers, each contributing two samples \cite{mak2021}. Additionally, Bi et al.\cite{bi2019} and Tan et al. \cite{tan2016} also employed this dataset for gender prediction, also using the RDF dataset, in Chinese, which includes about 11,000 writers, each providing one sample. 
Other researchers used the offline QUWI dataset, which includes 1000 writers and four samples per writer, written in Arabic and English \cite{mirz2016,gat2020,akba2016,gat2018}.

Few studies have been made on gender classification using online handwriting dataset :
Cordasco et al. utilized an online handwriting dataset for gender classification, consisting of approximately 250 adult males and females \cite{corda2020}. 
Erbilek et al. also used an online handwriting dataset comprising 100 participants \cite{erbi2016}.

Studies on gender prediction have predominantly focused on adults or young adults rather than children.

Research has been conducted on linking academic success with handwriting \cite{car2017,son2006}, but no research has yet been conducted on predicting academic performance from handwriting.

\section{Wacom Dataset} %Adrian [2 page]

\subsection{Overview} %Nishizawa [1 pages]

\begin{figure}[b]
    \centering
    \includegraphics[width=0.6\textwidth]{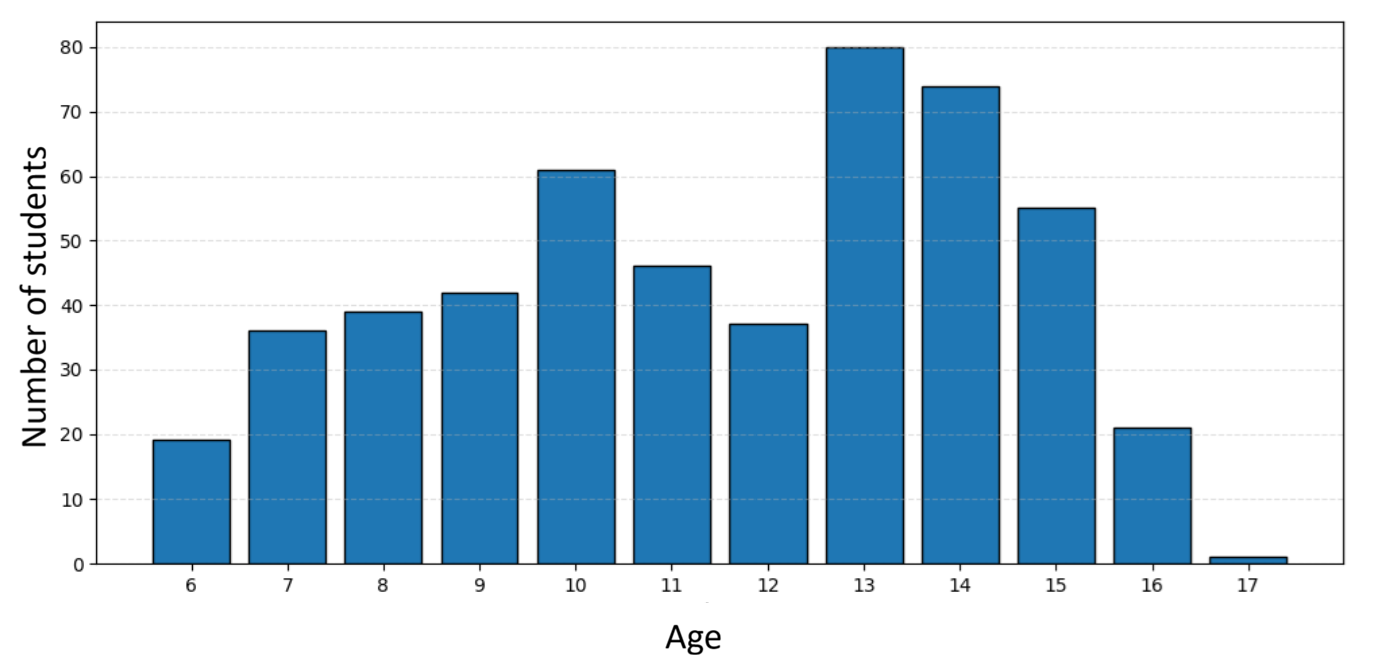}
    \caption{Distribution of the ages of the students in Wacom dataset.}\label{age}
\end{figure}

The Wacom dataset 
%-- added by kise
\footnote{This dataset was collected under the approval of the Ethical Review Committee of Tokyo University of Agriculture and Technology (240205-0411) with the consent of the respondents and their parents.}
%---
is an online dataset composed of numerous drills collected from Japanese students starting from the first year of elementary school (6 years old) to last year of junior high school (15 years old) which are 9 grades in total. In this paper, those grades are referred to as follows : for elementary school, the grades are referred as grade 1 to grade 6, and grade 7 to 9 for junior high school. The age of a student may differ from another one in the same grade. The distribution of the ages of the students is shown in Figure~\ref{age}. For grades 1 to 4, the students answer drills from 2 different school subjects : Japanese language and mathematics. For grades 5 and 6, they answer drills from 3 subjects : Japanese language, mathematics and English. Finally, for grade 7 to 9, they answer drills from 6 different subjects : Japanese language, mathematics, writing of kanjis (Japanese characters), English vocabulary, English, and calculations. 

Around 50 students can be found in each grade and each student answered around 110 drills per school subject in her/his grade for a total of around 145 000 drills, which makes the Wacom dataset one of the largest online handwriting datasets. The distribution of boys and girls can be found in Table~\ref{tab_gender_distribution}.

Each data sample includes the following information : the student's grade and age, gender, writing hand, and dominant hand. The handwriting of various students is recorded at a frequency of 480 Hz. Each data point contains the following information: the timestamp of the point relative to timestamp 0, the $x$, $y$, and $z$ coordinates of the pen relative to the device, the pressure of the pen, the tilt of the pen on the $x$ and $y$ axes, and the width of the pen tip. These points are grouped into strokes, which commence when the pen is in the “down” state and conclude when the pen is lifted.

\begin{table}[t]
\caption{Distribution of students across grades.}
\label{tab_gender_distribution}
\centering
\begin{tabular}{|c|c|c|c|}
\hline
\textbf{Grade} & \textbf{Boys} & \textbf{Girls} & \textbf{Total} \\
\hline
1 & 27 & 23 & 50 \\
\hline
2 & 24 & 26 & 50 \\
\hline
3 & 23 & 27 & 50 \\
\hline
4 & 17 & 30 & 47 \\
\hline
5 & 22 & 28 & 50 \\
\hline
6 & 30 & 20 & 50 \\
\hline
7 & 27 & 38 & 65 \\
\hline
8 & 24 & 33 & 57 \\
\hline
9 & 29 & 37 & 66 \\
\hline
\textbf{All} & \textbf{223} & \textbf{262} & \textbf{485} \\
\hline
\end{tabular}
\end{table}

\subsection{Subset of the Dataset : Japanese Language} %Nishizawa [1 pages]

\begin{figure}[htbp]
    \centering

    \begin{subfigure}{0.9\textwidth}
        \centering
        \includegraphics[width=\linewidth]{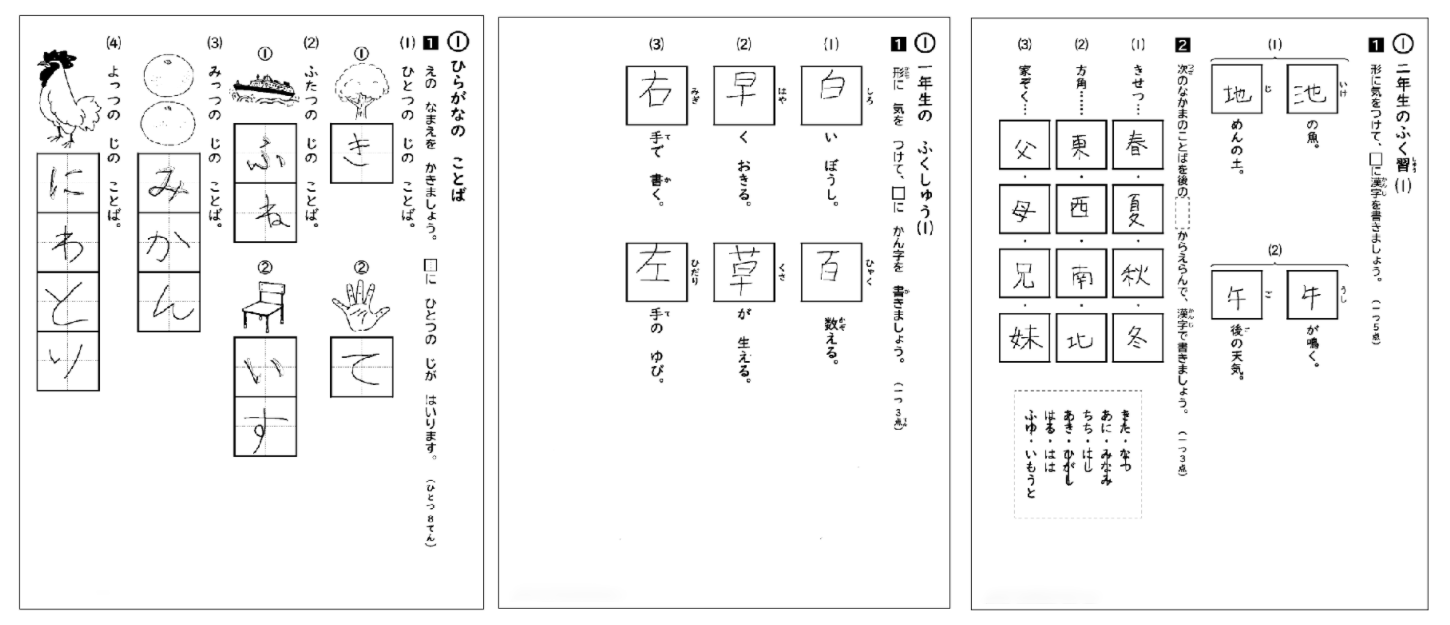}
        \caption{Grade 1-3}
        \label{fig:123}
    \end{subfigure}
    \hfill
    \begin{subfigure}{0.9\textwidth}
        \centering
        \includegraphics[width=\linewidth]{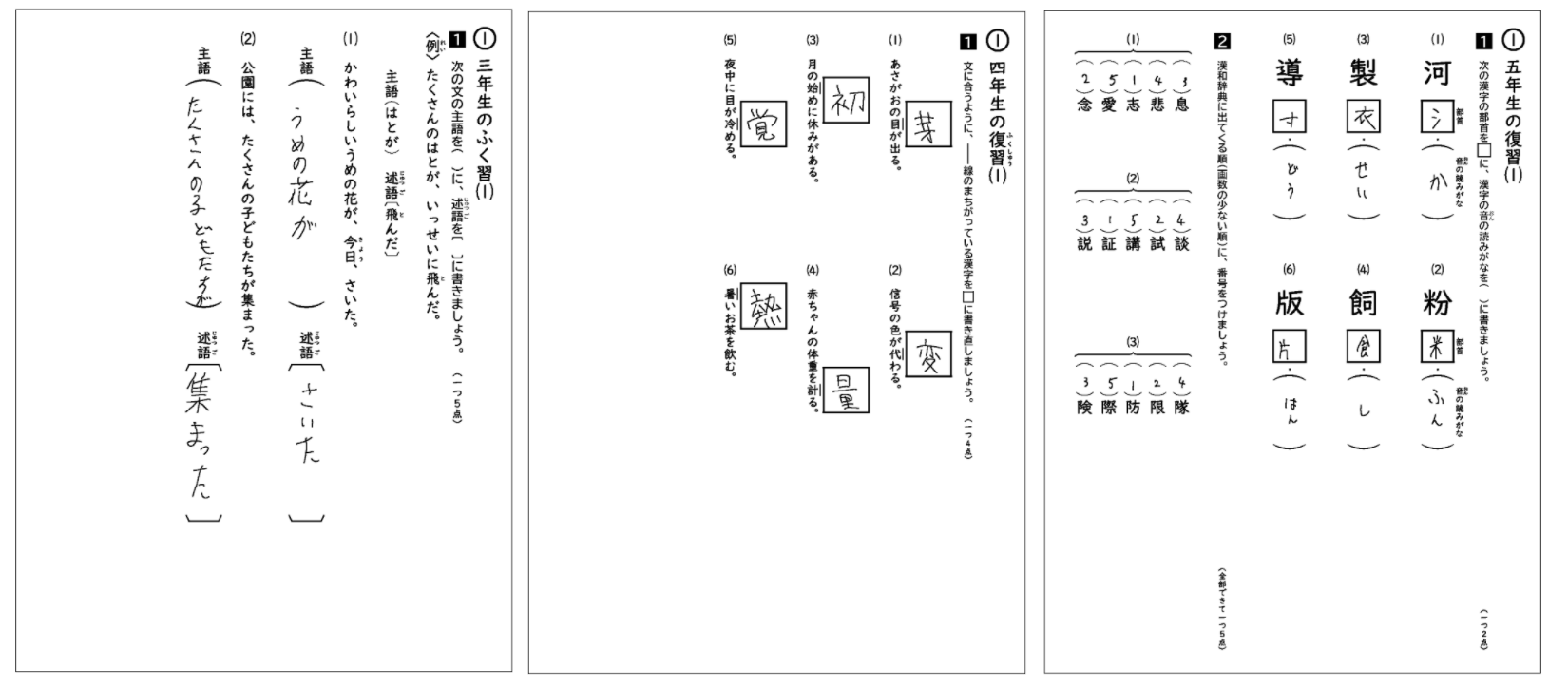}
        \caption{Grade 4-6}
        \label{fig:456}
    \end{subfigure}
    \hfill
    \begin{subfigure}{0.9\textwidth}
        \centering
        \includegraphics[width=\linewidth]{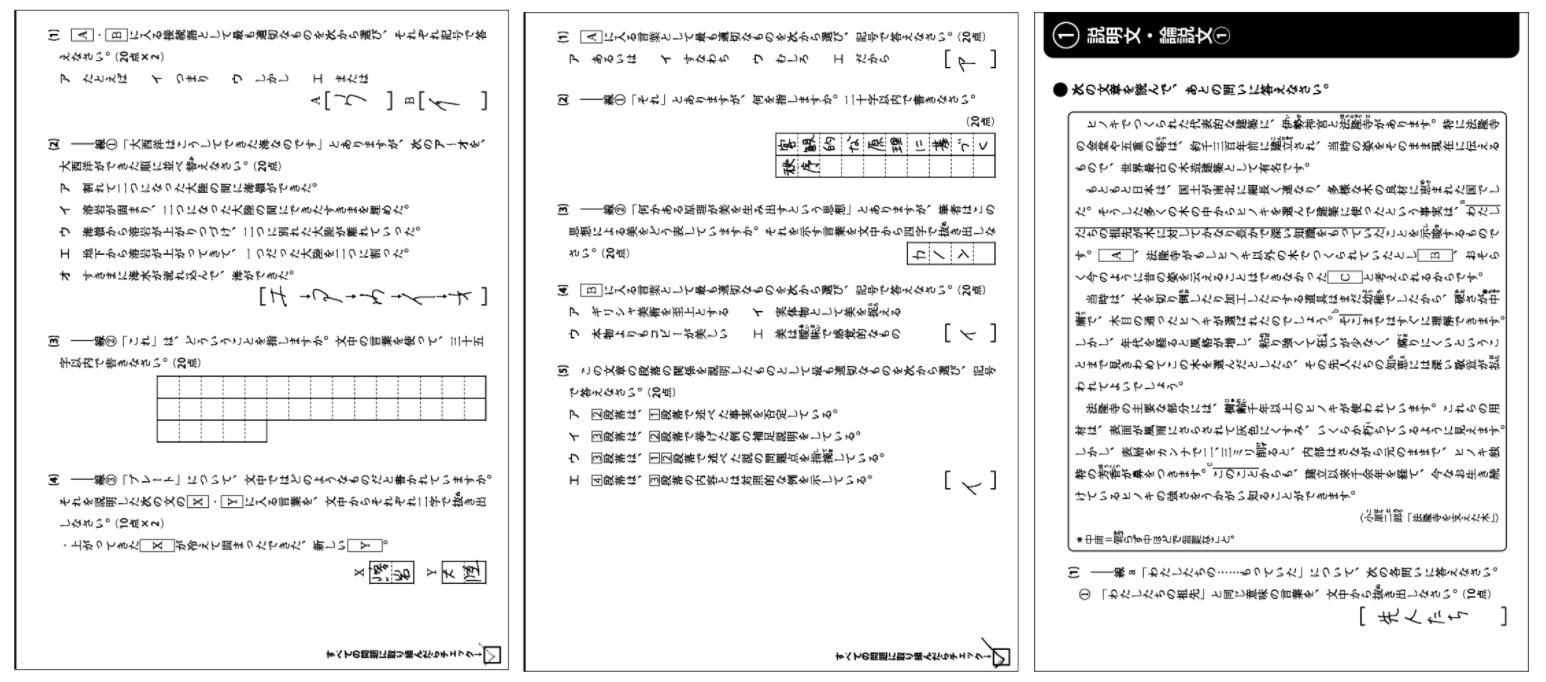}
        \caption{Grade 7-9}
        \label{fig:789}
    \end{subfigure}
    \caption{Examples of drills from subject “Japanese language” for grade 1 to 9.}
    \label{fig1}
\end{figure}

\begin{figure}[b]
    \centering
    \includegraphics[width=0.6\textwidth]{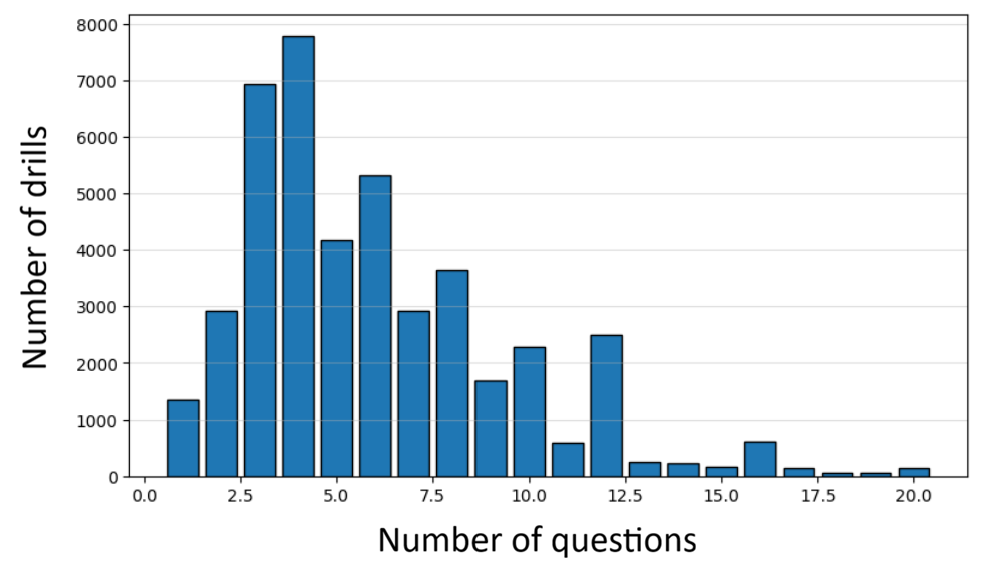}
    \caption{Distribution of number of questions in the drills of the subject Japanese language.}\label{questions}
\end{figure}

\begin{figure}[htbp]
    \centering

    \begin{subfigure}[t]{0.49\textwidth}
        \centering
        \includegraphics[width=\linewidth]{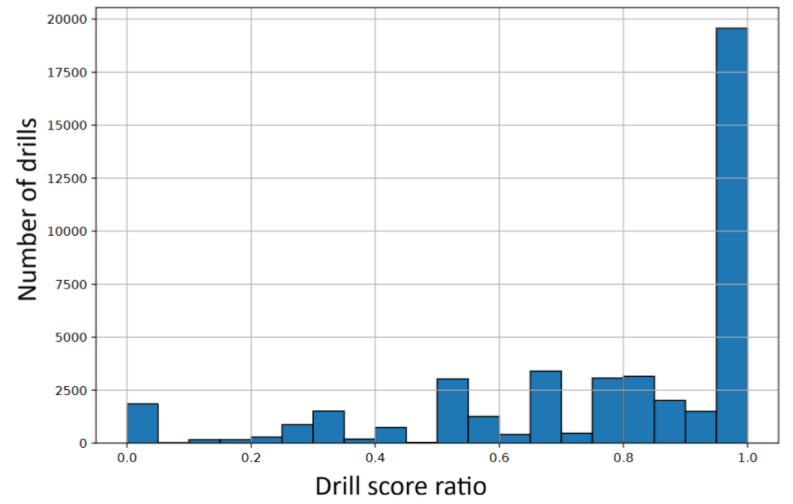}
        \caption{Distribution of the score ratio per drill.}
        \label{fig:ex_histo}
    \end{subfigure}
    \hfill
    \begin{subfigure}[t]{0.49\textwidth}
        \centering
        \includegraphics[width=\linewidth]{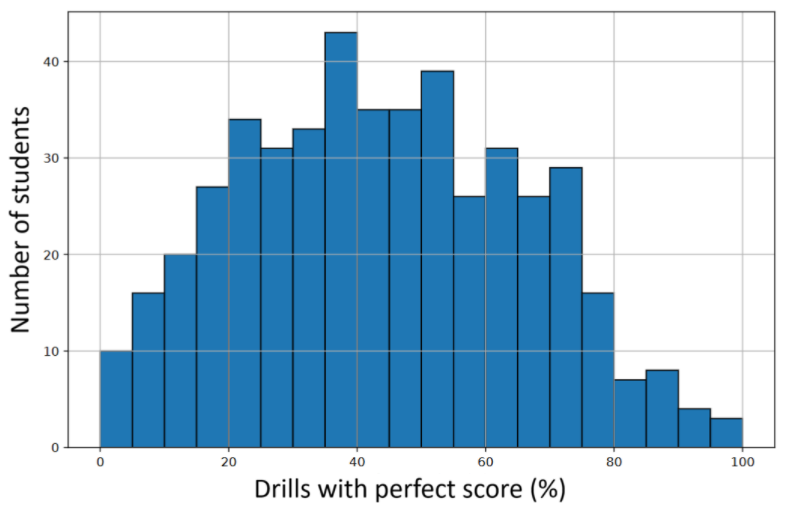}

        \caption{Percentage of perfectly answered drills per student.}
        \label{fig:perfect_ratio}
    \end{subfigure}

    \caption{Score distributions analysis in Wacome dataset}
    \label{score_fig}
\end{figure}

Figure~\ref{fig1} shows examples of drills from subject Japanese language. Each drill comprises 1 to 20 questions, and is then marked by humans. The distribution of the number of questions is shown in Figure~\ref{questions}. A correct response is awarded a score of “1” whereas an incorrect response receives a score of “0”. The score of a drill is a sum of all the scores of questions included in it. We calculate the score ratio by dividing the number of correct answers by the number of questions in the drill. The distribution of the score can be seen in Figure~\ref{score_fig}. As shown in Figure~\ref{score_fig}(a), the distribution of the score ratio is highly unbalanced, with around half of the drills marked as a perfect score. Figure~\ref{score_fig}(b) shows that few students perfectly answered all the drills, while most students perfectly answered to 40 to 50\% of the drills.

The numerous available features in the dataset open the possibility for new tasks or tasks that have not been extensively explored, such as estimating a student's academic performance.

\section{Prediction and Classification Methods} 

\subsection{Features} %1 page

Using the Wacom dataset, we extract three types of features.

Basic features are derived by calculating statistical measures such as the mean and standard deviation for parameters including speed, pressure, and tilt along both the $x$ and $y$ axes. Speed is specifically computed by measuring the Euclidean distance between two consecutive points within the same stroke, then dividing this distance by the time difference between their corresponding timestamps. This approach captures the dynamic movement characteristics of the stroke with respect to both spatial displacement and temporal progression. By extracting these features, the analysis captures not only the central tendency (mean) but also the variability (standard deviation) of the pen’s behavior during writing or drawing.

We also extract entropy features to capture the variability and distribution patterns within the metrics recorded for each drill using Shannon entropy as in \cite{rosso2016}. Entropy, in this context, quantifies the degree of randomness or disorder in the data, reflecting how the values of acceleration, pressure, and tilt are spread or concentrated over time during the drill. 

The specific features computed from these metrics include the mean and standard deviation of acceleration, pressure, and tilt across all drills of each student, which provide summary statistics of central tendency and dispersion. These statistical measures, combined with entropy, offer a comprehensive characterization of the drill’s dynamics.

%\begin{equation}
%p_i = \frac{n_i}{\sum_{j=1}^{N} n_j}
%\end{equation}

%\begin{equation}
%H = - \sum_{i=1}^{N} p_i \ln(p_i)
%\end{equation}

\begin{equation}
H_{\mathrm{norm}} = \frac{- \sum_{i=1}^{N} p_i \ln(p_i)}{\ln(N)}
\end{equation}
where $N$ is the number of histogram bins, $n_i$ is the number of samples in bin $i$, $p_i$ is the empirical probability of bin $i$.\\

The features extracted from the sigma-lognormal model provide a detailed characterization of handwriting by modeling the velocity profiles of individual pen strokes. Parameter extraction was carried out using the implementation by Lai et al.\cite{lai2022synsig2vec}\footnote{\url{https://github.com/LaiSongxuan/SynSig2Vec}}, which is derived from the method developed by O'Reilly and Plamondon\cite{oreilly2009sigma}. 
%--- commented out by kise
%The extraction method guarantees a signal-to-noise ratio of at least 15 dB based on velocity, which is typically deemed adequate for kinematic studies. 
%%%--------- Newly added part
On the Japanese language subset of the Wacom dataset, we achieved a mean signal-to-noise ratio (SNR) of $27.1 \pm 6.0$~dB, indicating good reconstruction quality for the sigma-lognormal model.
%-----
These features effectively capture the dynamic properties of handwriting movements, such as stroke timing, amplitude, and shape, which are directly influenced by the neuromotor control mechanisms governing fine motor skills. In our task we concentrate on 7 features, which are described in Table~\ref{tab_lognormal_features}.

\begin{table}[t]
\caption{Description of Lognormal Stroke Features}
\label{tab_lognormal_features}
\centering
\begin{tabular}{|p{4cm}|p{8cm}|}
\hline
\textbf{Feature} & \textbf{Description} \\
\hline

Number of Lognormal Components ($C$)
& The number of lognormal components required to model each stroke. 
This parameter reflects the degree of motor segmentation. \\
\hline

Velocity Profile Signal-to-Noise Ratio (SNR)
& Signal-to-noise ratio of the velocity profile, which quantifies how 
well the reconstructed lognormal model explains the observed 
kinematic signal. \\
\hline

Normalized SNR (SNR/$C$)
& Signal-to-noise ratio divided by the number of lognormal components, 
normalizing reconstruction quality by model complexity. \\
\hline

Time Dispersion ($D$)
& The lognormal time dispersion parameter controlling the temporal 
spread of each motor impulse. It is associated with the duration 
and fluidity of the movement. \\
\hline

Time Delay ($t_0$)
& The onset timing of motor commands within the stroke sequence. \\
\hline

Location Parameter ($\mu$)
& Determines the temporal positioning of each motor impulse within 
the stroke. \\
\hline

Scale Parameter ($\sigma$)
& Quantifies the temporal dispersion of the motor impulse on a 
logarithmic scale. \\
\hline

\end{tabular}
\end{table}

For basic features and entropy features, we first compute those features drill wise, and then summarize them at a student level by computing the mean and standard deviation on those features for all the drills of a student for a specific subject. For features extracted with the sigma-lognormal model, we first extract the mean and standard deviation of each feature in each drill, and then aggregate them at a student level by computing the mean of the extracted features.

\subsection{Prediction and Classification Method} %Adrian [2 pages]

For the prediction and classification, we employ both Linear or Logistic Regression models and Random Forest algorithms.

In the Linear or Logistic Regression models, multicollinearity among features is addressed through the calculation of the Variance Inflation Factor (VIF) during each fold of the cross-validation on the training set. Features exhibiting high collinearity (VIF above 4) are systematically removed to enhance model reliability. Following this, the Akaike Information Criterion (AIC) \cite{aka2025} is computed on the training sets to assess model quality, balancing performance with model complexity. This iterative process of feature elimination and AIC evaluation continues until the AIC values converge.

\subsubsection{Grade Prediction} %Nishizawa [1 pages]

The task of grade prediction treats student grades as ordinal variables on a scale from 1 to 9. This approach allows the model to capture the relative ranking and progression between grades, which is better for more nuanced and meaningful predictions.  

\subsubsection{Gender Classification} %Adrian [2 pages]
The task of gender classification is framed as a binary classification problem, where the goal is to accurately predict whether a student is “male” or “female” based on the available features.

\subsubsection{Academic Performance Classification}

The objective of academic performance classification is to determine whether a student achieves more than 45\% of drills with perfect scores. We selected this task due to the imbalance in the scores of the students fo each drill, with around half of the drills being answered perfectly, as shown in Figure~\ref{score_fig}(a). The 45\% threshold was selected to ensure a balanced distribution between students whose ratio of perfect scores exceeds the threshold and those whose ratio falls below it, with approximately 53\% of students having a ratio of perfect scores under 45\%.

\section{Experimental Results and Discussion}

\subsection{Methodology}
For the prediction and classification, we only focus on the subject “Japanese language”, which is the only common school subject that is mostly composed of handwritten characters and not numbers.

The Random Forest configuration used for all the predictions and classification includes 1200 estimators to enhance stability and reduce variance in the predictions, while setting the minimum leaf size to 2 helps prevent overfitting by ensuring that terminal nodes contain a sufficient number of samples. Notably, no maximum depth limit is imposed on the trees, allowing them to grow fully and capture complex patterns in the data without artificial constraints.

The deliberate decision to avoid extensive hyperparameter tuning reflects a focus on comparative analysis of feature sets rather than on optimizing the model’s absolute performance. By using standard and conservative parameter settings, we aim to minimize the risk that observed performance differences arise from overfitting or tuning artifacts, thereby improving the reliability of the results. 

For the evaluation of the models, a 5-fold cross-validation approach is employed, with splits performed at a student level : we divide the dataset into 5 folds of same size, each fold representing a small portion of the data with as much balance in the classes as possible.

For grade, predictions are evaluated using the explained variance $R^2$, which measures the proportion of the variance in the observed data that is explained by the model and Root Mean Square Error (RMSE), which denotes the root mean square error measuring the average magnitude of prediction errors.

%\begin{equation}
%R^2 = 1 - \frac{\sum_{i=1}^{N} (y_i - \hat{y}_i)^2}{\sum_{i=1}^{N} %(y_i - \bar{y})^2}
%\end{equation}

%\begin{equation}
%\mathrm{RMSE} = \sqrt{\frac{1}{N}\sum_{i=1}^{N} (y_i - \hat{y}_i)^2}
%\end{equation}

%where:
%\begin{itemize}
%    \item $N$ is the number of observations,
%    \item $y_i$ is the observed (actual) value of sample $i$,
%    \item $\hat{y}_i$ is the predicted value of sample $i$,
%    \item $\bar{y}$ is the mean of the observed values,

%\end{itemize}

For gender classification and academic performance classification, the results are benchmarked against a simple baseline method. The baseline consists in only predicting the dominant class.

\subsection{Results of Grade Prediction} %Nishizawa [2 pages]

\begin{figure}[htbp]
\centering

\begin{subfigure}[t]{0.9\textwidth}
    \centering
    \includegraphics[width=\linewidth]{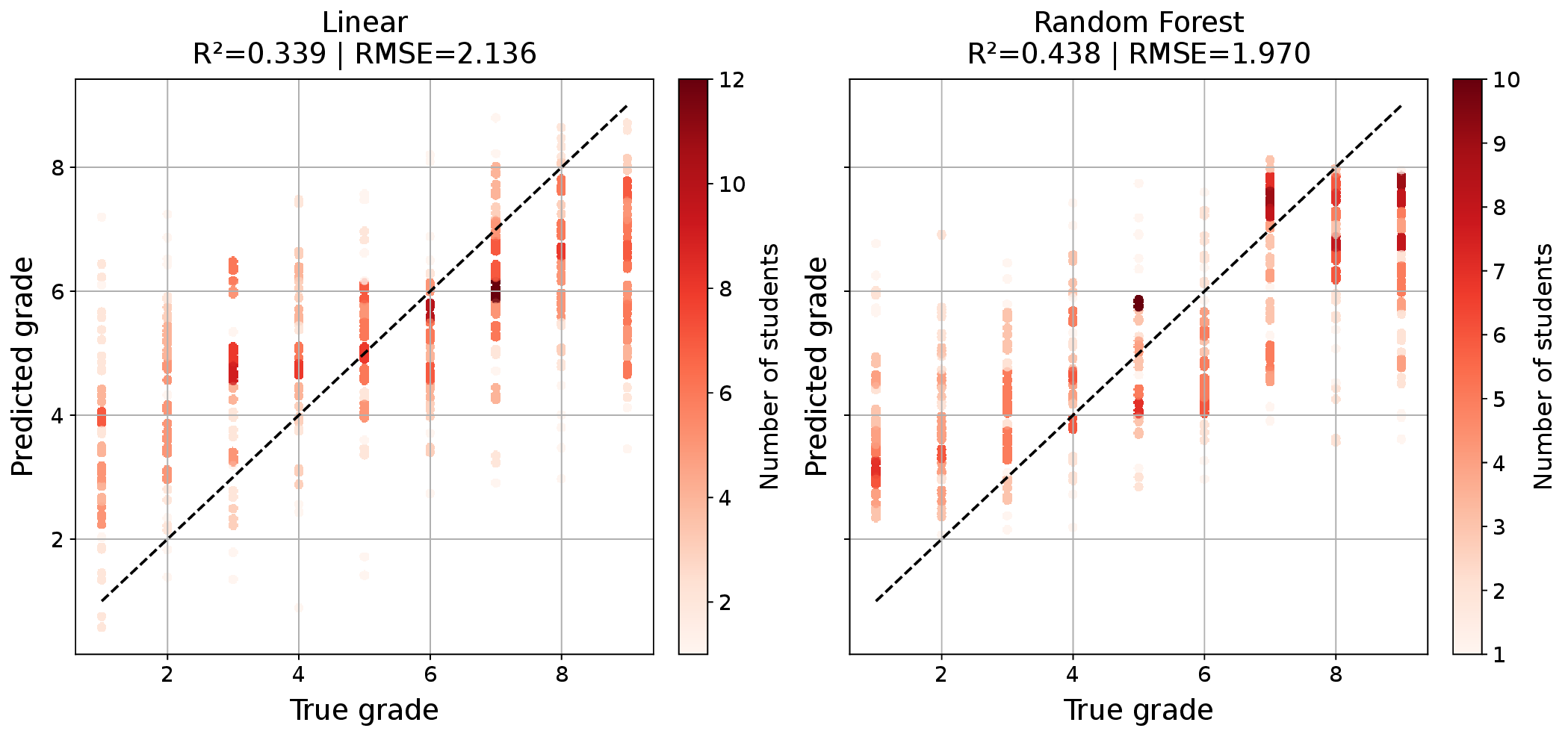}
    \caption{With base features}
\end{subfigure}
\hfill
\begin{subfigure}[t]{0.9\textwidth}
    \centering
    \includegraphics[width=\linewidth]{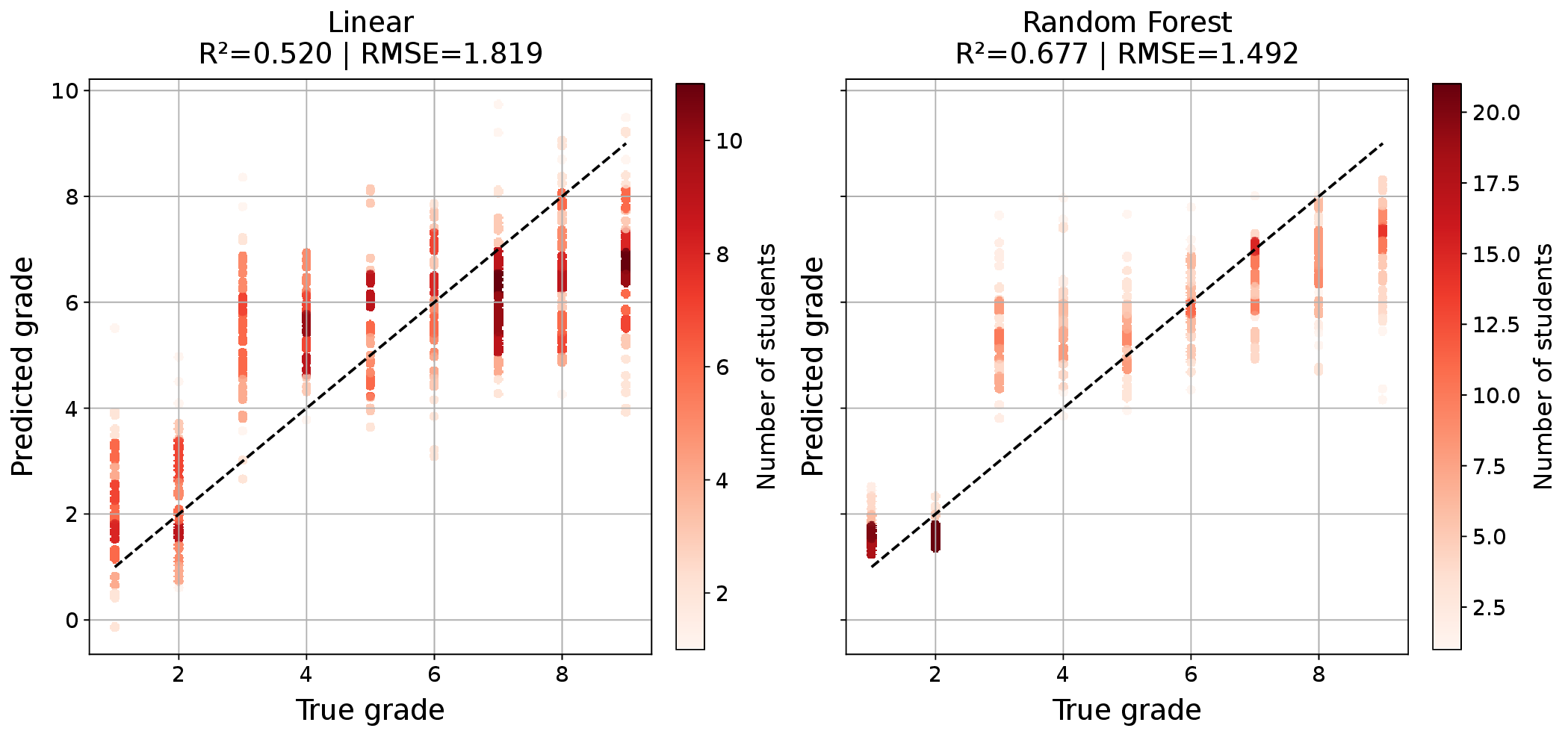}
    \caption{With entropy features}
\end{subfigure}
\hfill
\begin{subfigure}[t]{0.9\textwidth}
    \centering
    \includegraphics[width=\linewidth]{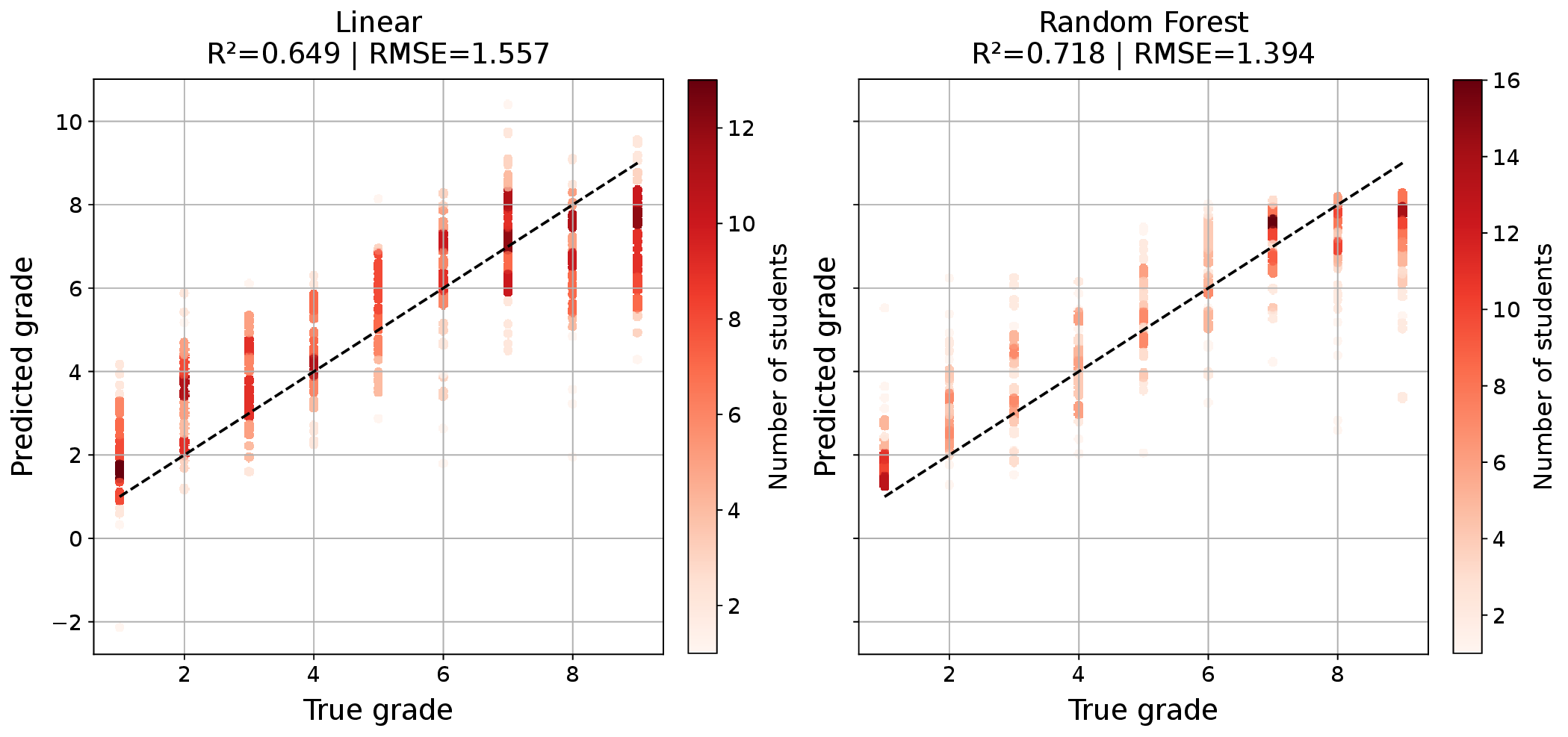}
    \caption{With sigma-lognormal features}
\end{subfigure}

\caption{Prediction of the grade for each set of feature using Linear Regression (left) and Random Forest (right) using 5-fold cross validation. One point corresponds to a prediction of a point from the test fold.}
\label{fig_grade}
\end{figure}

Figure~\ref{fig_grade} shows the results of the grade prediction. The results indicate that the features contain information pertinent to the students' grades. For the basic set of features, the models exhibit the least favorable performance. Examination of Figure~\ref{fig_grade} reveals that the model predominantly distinguishes between two classes: between grades 1 to 6 (elementary school), where the model exhibits limited predictive capacity to differentiate among these grades, and similarly for grades 7 to 9 (junior high school). However, the model appears to effectively differentiate between these two groups.

For the entropy features, the $R^2$ value is higher for the Random Forest, indicating significantly improved performance compared to the basic features. The entropy features reveal three distinct clusters: the model accurately differentiates the first two grades (1 and 2) from the other grades and can also generally separate grades of junior high school students (grades 7 to 9) from lower grades. However, it is unable to distinguish between grades within these clusters.

Finally, for the sigma-lognormal features, the Random Forest shows the best performance out of all models. The results show that with this set of features, the model performs better for predicting a student’s grade accurately from grade 1 to 6 (elementary school), but still shows low performance for distinguishing the grades within junior high school (grade 7 to 9).

\begin{figure}
\centering
\includegraphics[width=0.8\textwidth]{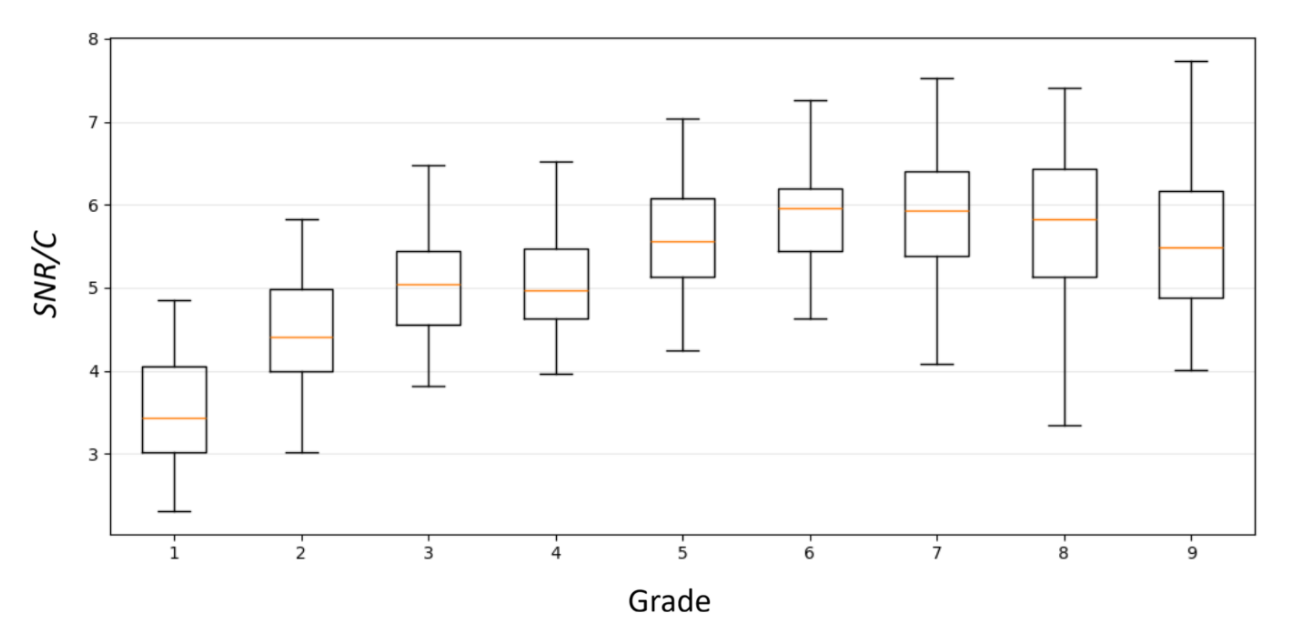}

\caption{Box plot of the signal-to-noise ratio over the number of lognormal components across all grades.} \label{snrc}
\end{figure}

Figure~\ref{snrc} describes the signal-to-noise ratio over the number of lognormal components across the grades, highlighting that young students' handwriting evolves toward a lognormal behavior.

\subsection{Results of Gender Classification} %Adrian [2.5 pages]
\begin{figure}[htbp]
\centering

\begin{subfigure}[t]{0.32\textwidth}
    \centering
    \includegraphics[width=\linewidth]{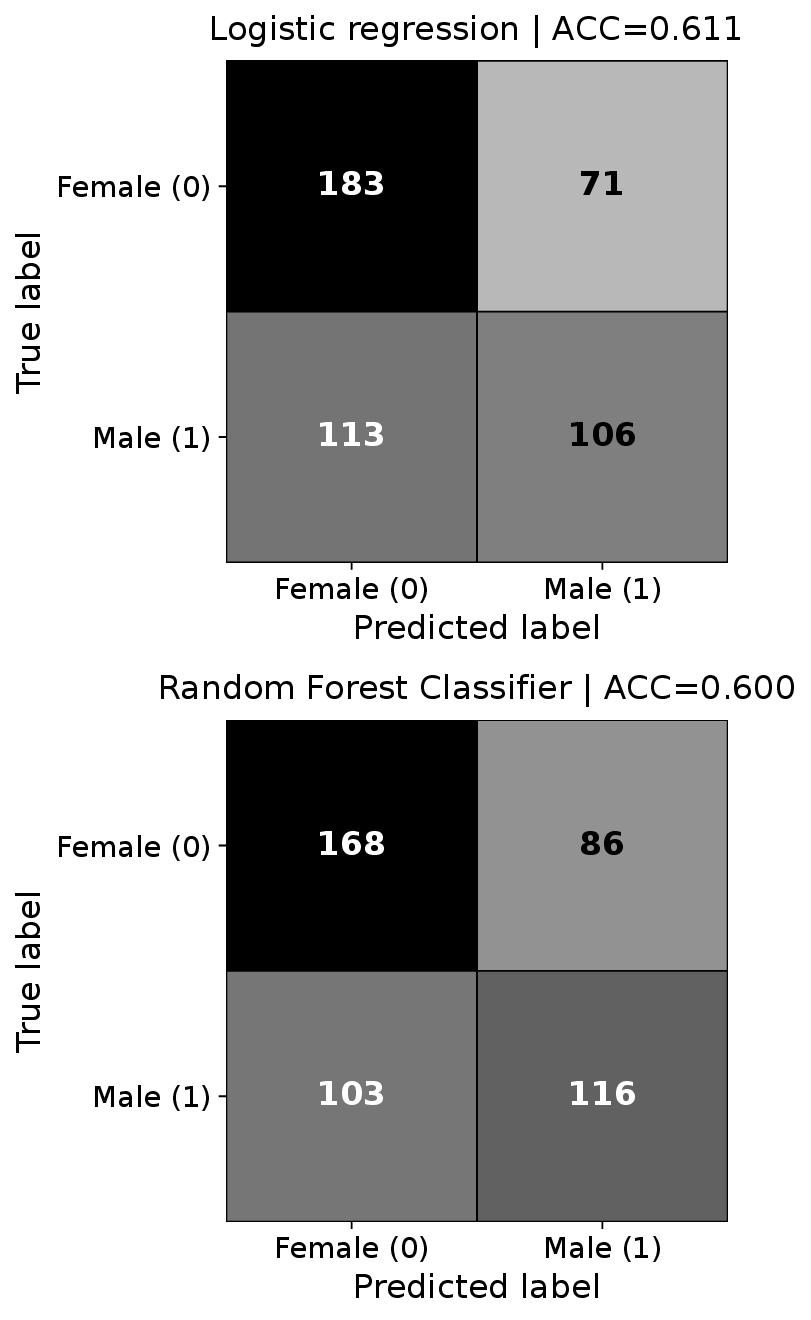}
    \caption{With base features}
\end{subfigure}
\hfill
\begin{subfigure}[t]{0.32\textwidth}
    \centering
    \includegraphics[width=\linewidth]{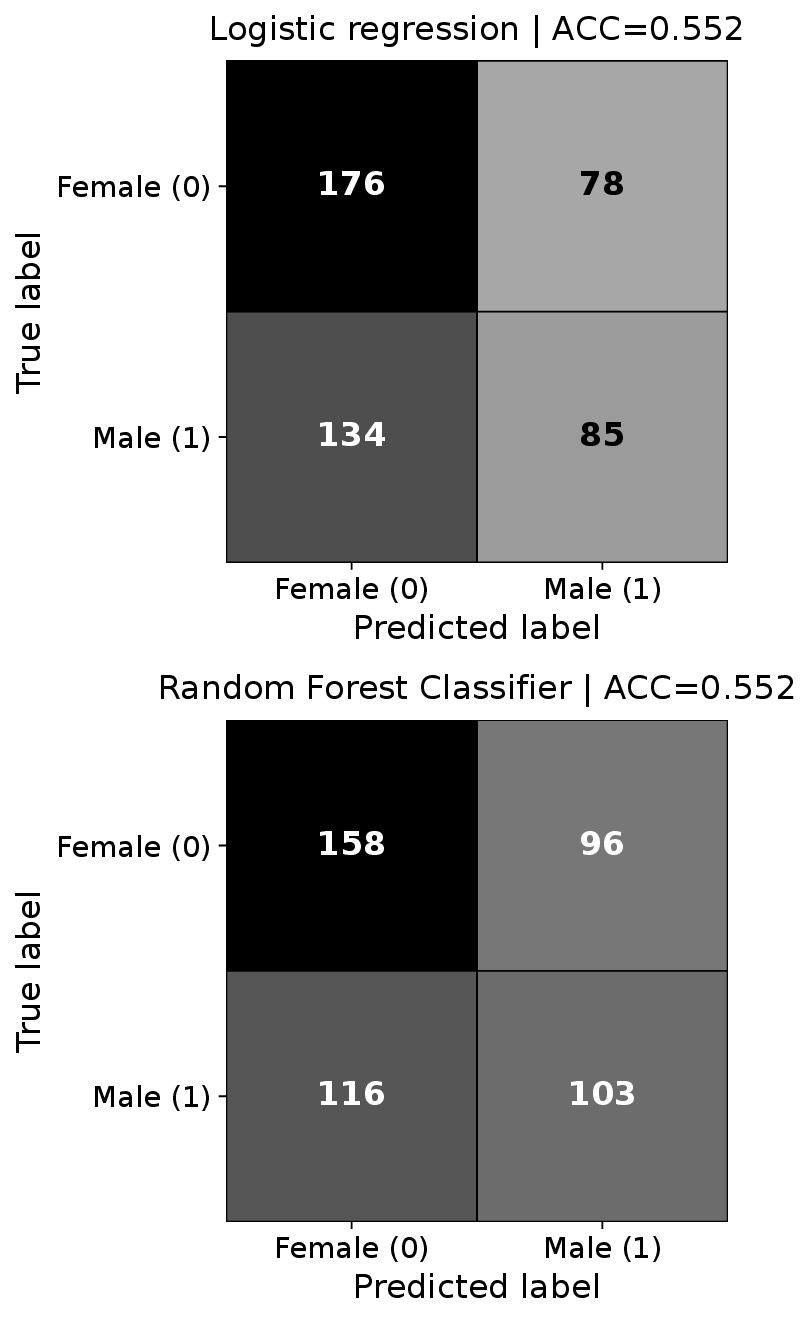}
    \caption{With entropy features}
\end{subfigure}
\hfill
\begin{subfigure}[t]{0.32\textwidth}
    \centering
    \includegraphics[width=\linewidth]{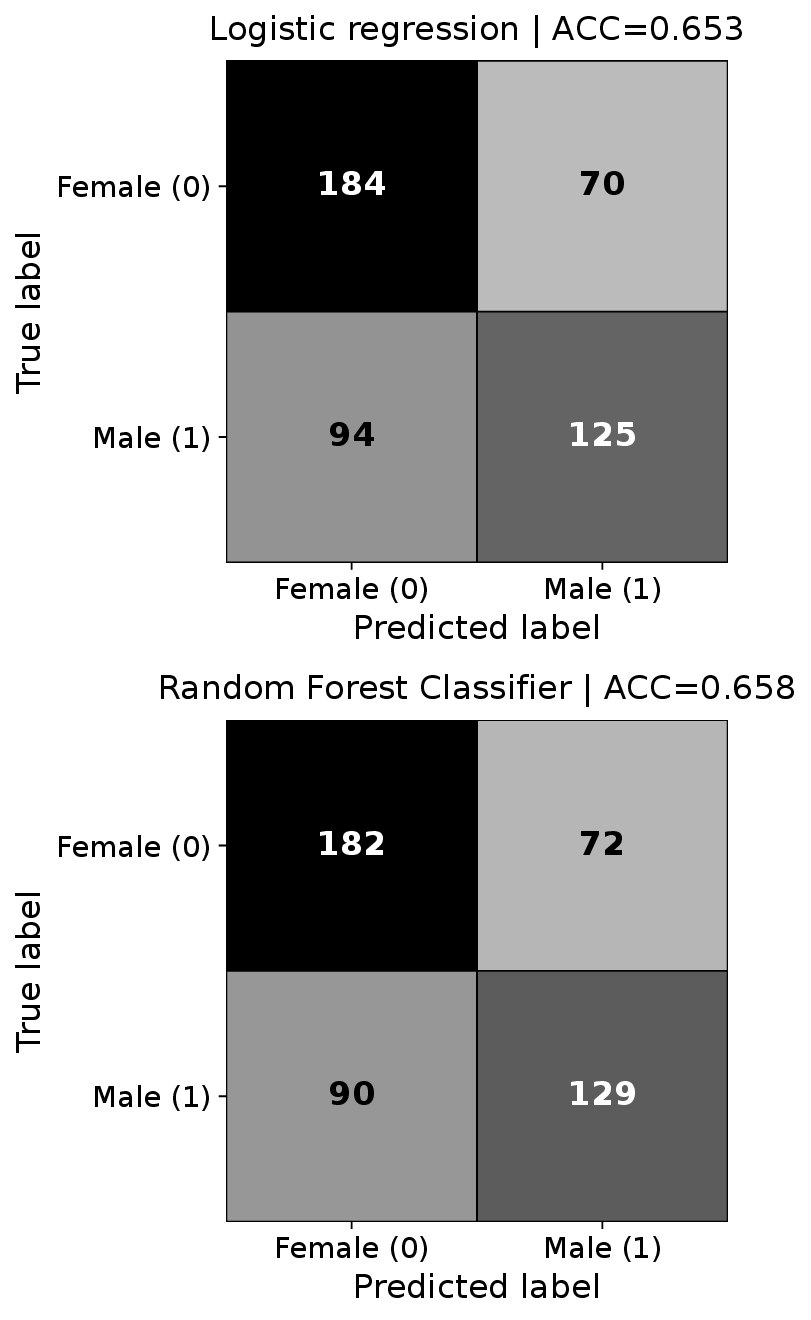}
    \caption{With sigma-lognormal features}
\end{subfigure}

\caption{Confusion matrix of the gender classification for each set of feature using Logistic Regression (top) and Random Forest (bottom) with 5-fold cross validation. } \label{fig_gender}
\end{figure}

Table~\ref{tab_gender} presents the performance of gender classification across all grades. The evaluation metrics include accuracy (ACC), F1-score (F1), and Area Under the ROC Curve (AUC). Figure~\ref{fig_gender} describes the confusion matrices of the classification task for each set of feature and each model.

The sigma-lognormal feature set again demonstrates superior performance across both classifiers. 

In contrast, the basic features exhibit lower performance. The F1-scores are lower, indicating a reduced balance between precision and recall. AUC values are also inferior to those obtained with sigma-lognormal features, indicating weaker overall discriminative signal. 

For this task, the entropy features perform the least effectively among the feature sets. Both classifiers achieve similarly low accuracy, the F1-scores remain relatively low, and AUC values are significantly lower than those obtained with sigma-lognormal features.

\begin{table}
\caption{Gender classification performance over all grades using 5-fold cross validation for all sets of feature compared to a majority-class baseline.}
\label{tab_gender}
\centering
\begin{tabular}{|l|cccc|cccc|c|}
\hline
\textbf{Feature Set} 
& \multicolumn{4}{c|}{\textbf{Logistic Regression}} 
& \multicolumn{4}{c|}{\textbf{Random Forest}} 
& \multicolumn{1}{c|}{\textbf{Majority-class baseline}} \\
\cline{2-10}
 & ACC & F1 & AUC &  
 & ACC & F1 & AUC &  
 & Majority ACC \\
\hline
Base
& 0.611 & 0.535 & 0.626 &
& 0.600 & 0.551 & 0.625 &
& 0.537 \\
\hline
Entropy  
& 0.552 & 0.445 & 0.574 &
& 0.552 & 0.493 & 0.595 &
& 0.537 \\
\hline
Sigma-lognormal       
& 0.653 & 0.604 & 0.725 &
& 0.658 & 0.614 & 0.693 &
& 0.537 \\
\hline
\end{tabular}
\end{table}

\subsection{Results of Academic Performance Classification} 

\begin{figure}[htbp]
\centering

\begin{subfigure}[t]{0.32\textwidth}
    \centering
    \includegraphics[width=\linewidth]{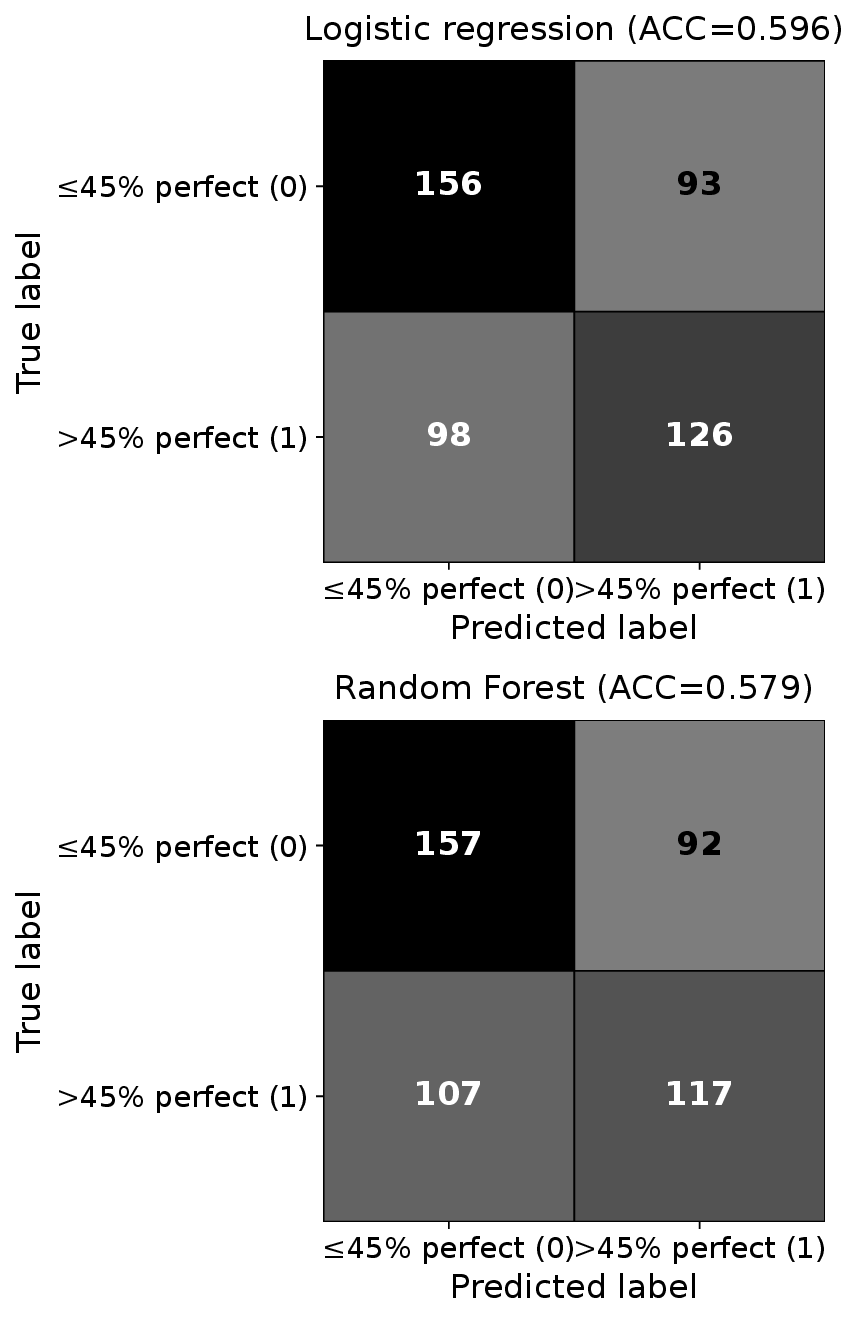}
    \caption{With base features}
\end{subfigure}
\hfill
\begin{subfigure}[t]{0.32\textwidth}
    \centering
    \includegraphics[width=\linewidth]{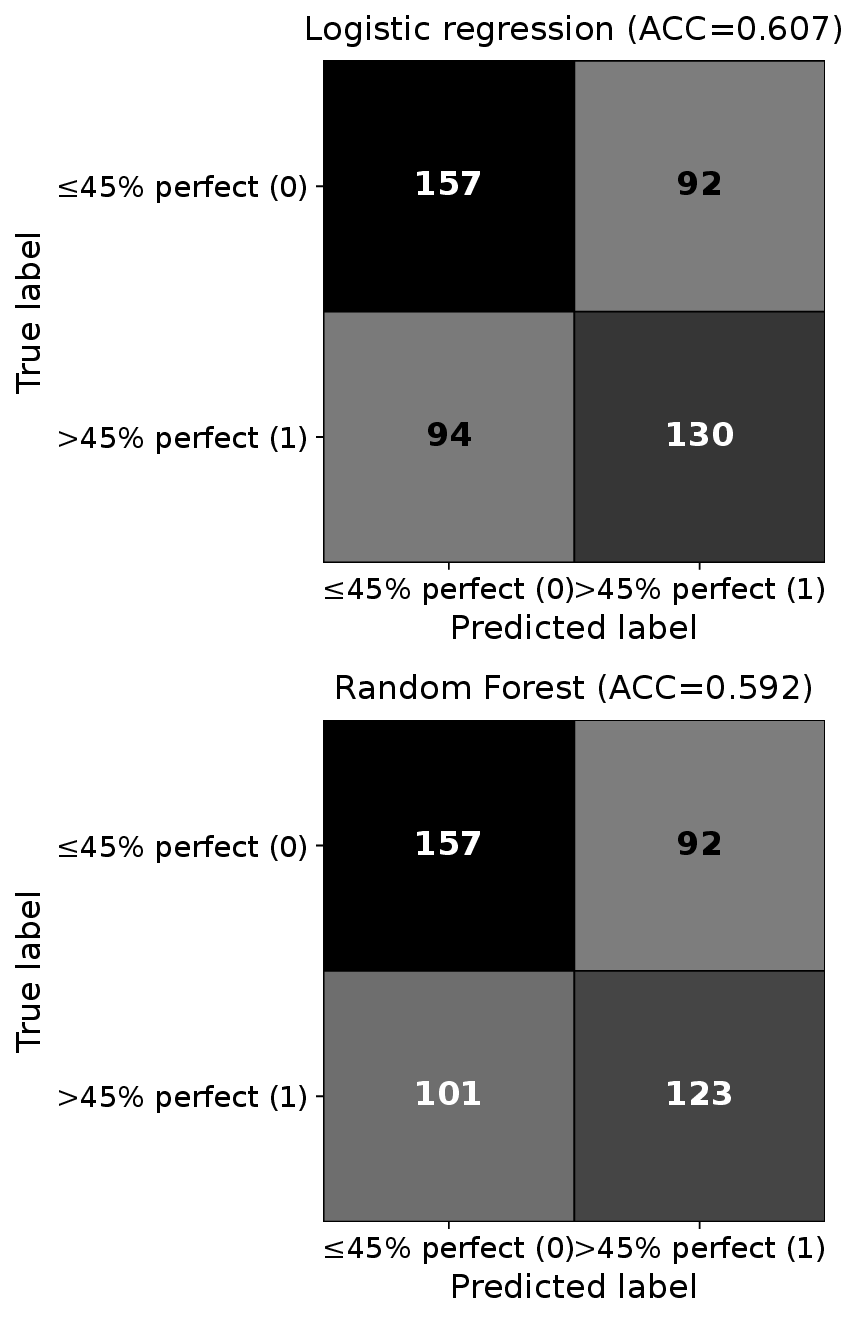}
    \caption{With entropy features}
\end{subfigure}
\hfill
\begin{subfigure}[t]{0.32\textwidth}
    \centering
    \includegraphics[width=\linewidth]{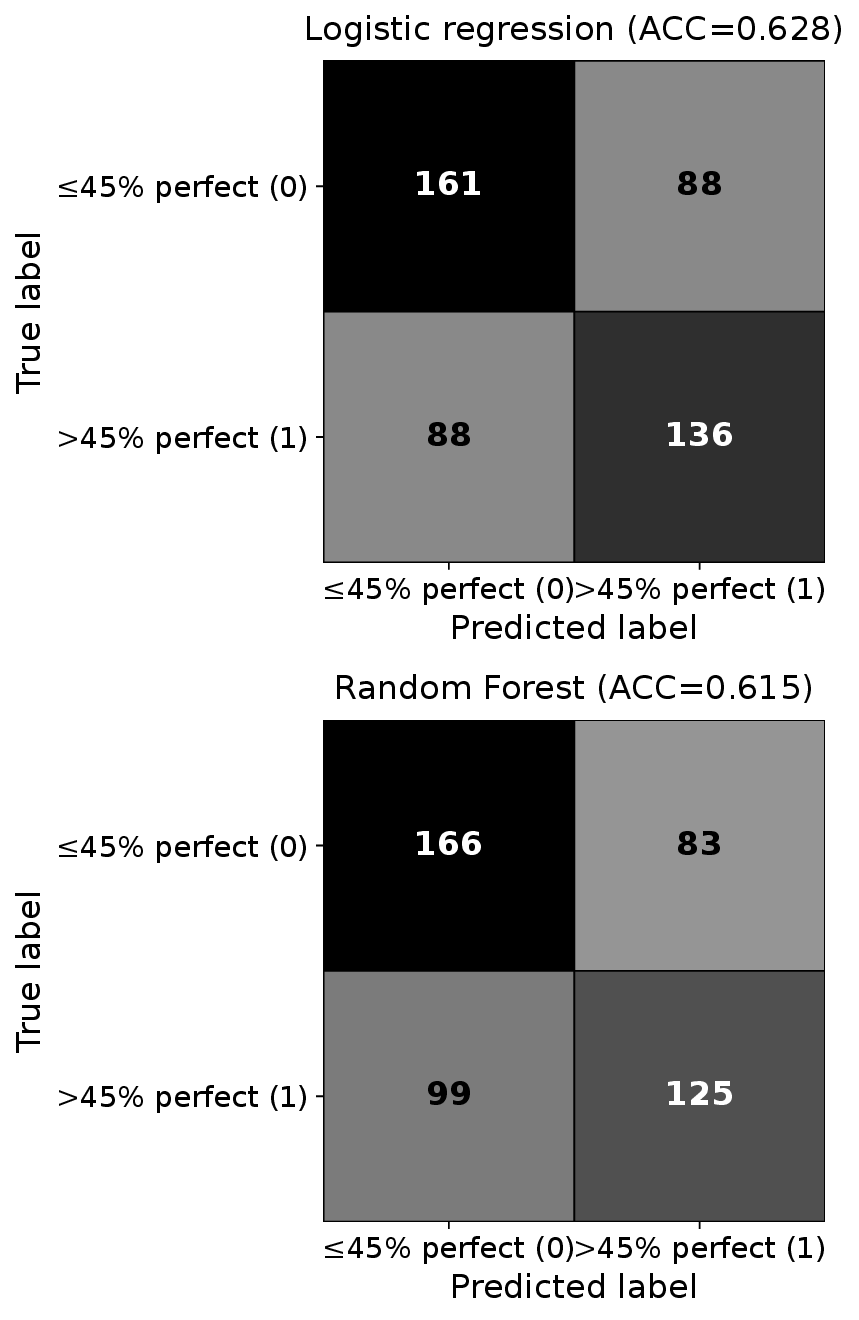}
    \caption{With sigma-lognormal features}
\end{subfigure}

\caption{Confusion matrix of the prediction of the academic performance of a student for each set of feature using Logistic Regression (top) and Random Forest (bottom) with 5-fold cross validation. } \label{fig_score}
\end{figure}

Table~\ref{tab_score} shows the classification outcomes at the student level for predicting whether a student surpasses 45\% of the ratio of perfectly answered drills. Figure~\ref{fig_score} describes the confusion matrices of the classification task for each set of feature and each model. The findings suggest that the features extracted from the sigma-lognormal model offer the most robust predictive signal among the three feature sets evaluated. Both Logistic Regression and Random Forest models exhibit the best performance when trained on the sigma-lognormal features. As shown in Table~\ref{tab_score}, AUC and F1-score consistently surpass those obtained with entropy features and basic features, indicating that the sigma-lognormal features encapsulate more task-relevant aspects of handwriting behavior pertinent to student performance. Although all models significantly outperform the majority baselines, the overall predictive efficacy remains moderate.

\begin{table}
\caption{Classification results over all grades of the overall academic performance of a student using 5-fold cross validation for all sets of feature compared to a “always majority” model.}
\label{tab_score}
\centering
\begin{tabular}{|l|cccc|cccc|c|}
\hline
\textbf{Feature Set} 
& \multicolumn{4}{c|}{\textbf{Logistic Regression}} 
& \multicolumn{4}{c|}{\textbf{Random Forest}} 
& \multicolumn{1}{c|}{\textbf{Baseline}} \\
\cline{2-10}
 & ACC & F1 & AUC &  
 & ACC & F1 & AUC &  
 & Majority ACC \\
\hline
Base
& 0.596 & 0.569 & 0.645 &
& 0.579 & 0.540 & 0.624 &
& 0.526 \\
\hline
Entropy
& 0.607 & 0.583 & 0.642 &
& 0.592 & 0.560 & 0.641 &
& 0.526 \\
\hline
Sigma-lognormal
& 0.628 & 0.607 & 0.673 &
& 0.615 & 0.579 & 0.665 &
& 0.526 \\
\hline
\end{tabular}
\end{table}

\subsection{Discussion} %Adrian [1.5 pages]

The results obtained in this study suggest that handwriting dynamics contain measurable information related to several student characteristics, including grade, gender, and academic performance. Across all tasks, the features extracted from the sigma-lognormal model achieved the best performance, indicating that features extracted from handwriting signals capture behavioral patterns that are not entirely random. However, the predictive performance remains moderate overall, highlighting both the potential and the current limitations of handwriting-based prediction.

For grade prediction, the models were able to capture broad differences between groups of students, particularly when considering larger grade intervals. The predictions indicate that handwriting characteristics evolve with schooling progression, although distinguishing between adjacent grades remains challenging. This difficulty may reflect the fact that handwriting development does not evolve linearly across all school years, and that variability between individuals can be comparable to differences between neighboring grades. While the obtained errors remain lower than those expected from random predictions, the results suggest that handwriting alone may not be sufficient to precisely determine fine-grained academic levels.

Gender classification produced modest improvements over the majority-class baseline. Although basic features and features extracted from the sigma-lognormal model allowed classifiers to identify patterns associated with gender above the chance level, the performance gap remains limited. This outcome suggests that any differences captured by handwriting dynamics are subtle and potentially influenced by multiple external factors. Consequently, these results should be interpreted cautiously, as handwriting behavior is shaped by educational context, personal habits, and individual variability rather than demographic characteristics alone.

Similarly, the classification of academic performance proved to be a demanding task. Academic success depends on numerous factors extending beyond motor execution, including cognitive abilities, motivation, and learning environment. Nevertheless, the fact that all models outperform baseline strategies suggests that certain aspects of writing behavior correlate with performance-related patterns, even if the relationship remains indirect.

Across the different experiments, the comparison between different features highlights the importance of feature representation. While performance varied depending on the task, no single representation provided near-perfect prediction, emphasizing that handwriting analysis captures only a limited portion of the variability associated with student characteristics. The observed improvements obtained with more structured feature representations suggest that incorporating temporal or dynamic properties of writing movements can be beneficial, although the precise mechanisms underlying these gains remain difficult to isolate within the current experimental framework.

Several methodological choices should also be considered when interpreting the results. The models were intentionally configured using conservative and consistent parameters in order to prioritize fair comparison between feature sets rather than maximizing absolute performance. While this approach strengthens the validity of comparative conclusions, it likely underestimates the achievable performance with extensive hyperparameter optimization or alternative learning strategies. In addition, aggregating handwriting features at the student level simplifies the prediction problem but may reduce sensitivity to short-term variations occurring across exercises or writing contexts.

\section{Conclusion} %Adrian [0.5]

This work explored the prediction and classification of student characteristics from handwriting dynamics using digital pen recordings. Unlike much of the existing literature, which primarily focuses on adults or clinical populations, this study considers school-aged children, whose handwriting reflects ongoing motor and cognitive development. This context introduces additional variability but also opens research perspectives, as understanding handwriting evolution during learning stages remains relatively underexplored.

Several limitations should be acknowledged. Although the dataset contains a large number of handwriting samples recorded at high temporal resolution, the present study relies exclusively on aggregated student-level features. While this choice simplifies modeling and enables direct comparison between feature representations, it likely discards valuable temporal information contained within individual strokes and drills. Future work could therefore investigate sequence-based or sample-level approaches capable of exploiting the richness of the raw signals.

Despite these limitations, the results demonstrate that handwriting dynamics contain meaningful signals related to student characteristics. In particular, features derived from structured representations of handwriting movements, including those obtained from the sigma-lognormal model, consistently showed promising performance across tasks, especially for grade prediction. These findings highlight the potential of kinematic handwriting analysis as a complementary approach for studying learning processes and developmental patterns and confirm the research conducted by Plamondon et al. \cite{plam2013}, suggesting that, over time, children’s handwriting tends toward an optimal lognormal motor organization. Further exploration using richer modeling frameworks and longitudinal analyses may help better leverage this potential in educational contexts.

\section*{Acknowledgments} %Adrian [0.5]

The authors would like to express their sincere gratitude to Mr. Toshihiko Horie and Mr. Takahiro Yamamoto of Wacom Co., Ltd. for providing the dataset used in this study. This research was supported by the JST research program CRONOS (Grant No. JPMJCS24K4).

%references [2 pages]

\bibliographystyle{splncs04}
\bibliography{ref}
\end{document}